\documentclass[epj]{svjour}

\usepackage{graphics}

\begin{document}
\title{A prototype industrial laser system for cold atom inertial sensing in space}
\subtitle{}
\author{
Romain Caldani\inst{1},
S\'{e}bastien Merlet\inst{1},
Franck Pereira Dos Santos\inst{1} \and
Guillaume Stern\inst{2},
Anne-Sophie Martin\inst{2},
Bruno Desruelle\inst{2},
Vincent M\'{e}noret\inst{2}\thanks{E-mail: vincent.menoret@muquans.com}
%\thanks{\emph{Present address:} Insert the address here if needed}%
}                      

\institute{
LNE-SYRTE, Observatoire de Paris, Universit\'{e} PSL, CNRS, Sorbonne Universit\'{e}, 61 avenue de l'Observatoire, 75014 Paris, France \and
MUQUANS, Institut d'Optique d'Aquitaine, rue Fran\c{c}ois Mitterrand, 33400 Talence, France
}

\date{Received: date / Revised version: date}
% The correct dates will be entered by Springer

\abstract{
We present the design, realization, characterization and testing of an industrial prototype of a laser system, which is based on frequency doubling of telecom lasers and features all key functionalities to drive a cold atom space gradiometer based on the architecture proposed in \cite{trimeche2019}. Testing was performed by implementing the laser system onto a ground based atomic sensor currently under development. The system reaches a Technology Readiness Level (TRL) of 4, corresponding to an operational validation in a controlled environment. The optical architecture of the system can be adapted to other space mission scenarios.
\PACS{
      {PACS-key}{discribing text of that key}   \and
      {PACS-key}{discribing text of that key}
     } % end of PACS codes
} %end of abstract

\authorrunning{R. Caldani et al.}

\maketitle
\section{Introduction}
\label{sec:intro}

Cold atom inertial sensors have demonstrated excellent performances in terms of sensitivity, stability and accuracy \cite{hu2013,Gillot2014,sorrentino2014,freier2016,savoie2018,karcher2018,asenbaum2017,overstreet2018}. In space, cold atom clouds in free-fall have a null velocity with respect to the science chamber, meaning that the interrogation time $2T$ is not limited by the size of the setup as it is in ground-based instruments, but rather by the residual expansion of the cloud due to its finite temperature. Sensitivity, which scales like $T^2$ in most cold atom inertial sensors, can therefore be greatly improved in microgravity environments \cite{barrett2016,becker2018,elliott2018}. This makes atomic sensors interesting alternatives to classical instruments for future space missions dedicated to fundamental physics or geodesy \cite{carraz2014,hogan2011,altschul2015,chiow2015,hogan2016,tino2019}.

Most demonstrations of high performance atomic inertial sensing are laboratory-based experiments that do not meet the high level of qualification and testing required for an operation in space. Over the last few years, significant efforts have been conducted to develop mobile instruments that can be operated outside the lab \cite{barrett2016,becker2018,bidel2018,menoret2018}. Each of these demonstrations required to improve the compacity and reliability of the laser system, which is a key subsystem of a cold atom sensor \cite{cheinet2006,schmidt2011,Merlet2014,leveque2014,schkolnik2016,cheng2017,zhang2018}. In the perspective of future space missions, laser systems will require further improvements, especially in terms of reliability and resistance to external conditions such as temperature, vibrations and radiations.

In this paper, we present the design, realization and characterization of a prototype industrial laser system (ILS) designed to meet the scientific goals of a space-based gravity gradiometry mission using cold Rubidium atoms \cite{trimeche2019,carraz2014}. The system is based on frequency doubling of telecom lasers operating at 1560~nm. This technology is capable of meeting both the scientific and operational specifications of the mission.

We first briefly recall the specifications of the laser based on the mission concept study, and describe the laser system focusing both on scientific performance and on technological choices that make it suited to a future space mission. We then show the results of optical characterizations. Finally, the laser was evaluated on a ground-based atomic gravity sensor.

\section{Laser system description}

\subsection{Mission concept and laser specifications}
This work uses as a baseline the Cold Atom Interferometer (CAI) gradiometry mission originally proposed in \cite{carraz2014}. In this space mission scenario, cold atom interferometers are used in a differential configuration to enable high sensitivity measurements of the Earth's gravity gradient. The specifications of the laser system were derived from that of the atom interferometry sequence \cite{trimeche2019}. Namely, it should feature several frequency-stabilized fibered outputs used for Rubidium cooling, detection, Bloch elevator and Raman interrogation (Table \ref{tab:specs}). In addition, the Raman output should have a narrow linewidth (10~kHz), and a high short-term power stability (0.1\% over 2~s).

\begin{table}
\caption{Main specifications for the laser system. Raman detuning is from the $|F'=1\rangle$ level.}
\label{tab:specs}       % Give a unique label
% For LaTeX tables use
\begin{tabular}{lccc}
\hline\noalign{\smallskip}
 & Cooling & Bloch & Raman \\
\noalign{\smallskip}\hline\noalign{\smallskip}
Wavelength (nm) & 780.24 & 780.24 & 780.24\\
Power (mW) & 200 & 200 & 45\\
Detuning (MHz) & $[-20 ; -120]$ & 100000 & -3400 \\
Linewidth (kHz) & $< 500$ & $<6000$ & $<10$ \\
Power stability (\%) & 1 & 1 & 0.1 in 2~s \\
\noalign{\smallskip}\hline
\end{tabular}
\end{table}

As proposed in \cite{trimeche2019}, we selected frequency-doubled telecom lasers to build the ILS. This approach relies on the use of laser diodes and amplifiers operating in the telecom C-band around 1560~nm and frequency-doubled to 780~nm in nonlinear crystals \cite{thompson03}. It benefits from the high maturity and reliability of telecom components, most of which are qualified according to the Telcordia standard. It has been used successfully in laboratory and mobile cold atom sensors over the last few years, meeting both optical and environmental constraints \cite{menoret2018,leveque2014,theron2017}. A significant advantage of this technology is the fact that it uses individual fibered components that are fusion spliced to one another to build the system. It is therefore convenient to replace a component by another when required and to include redundancy in the system by using fiber couplers to link the spare component to the rest of the system. Furthermore, this technology offers a significant level of space qualification as most of the optical components have either been qualified specifically or have qualified alternatives \cite{leveque2014}.

\subsection{Optical architecture}
\begin{figure}
\resizebox{0.48\textwidth}{!}{%
  \includegraphics{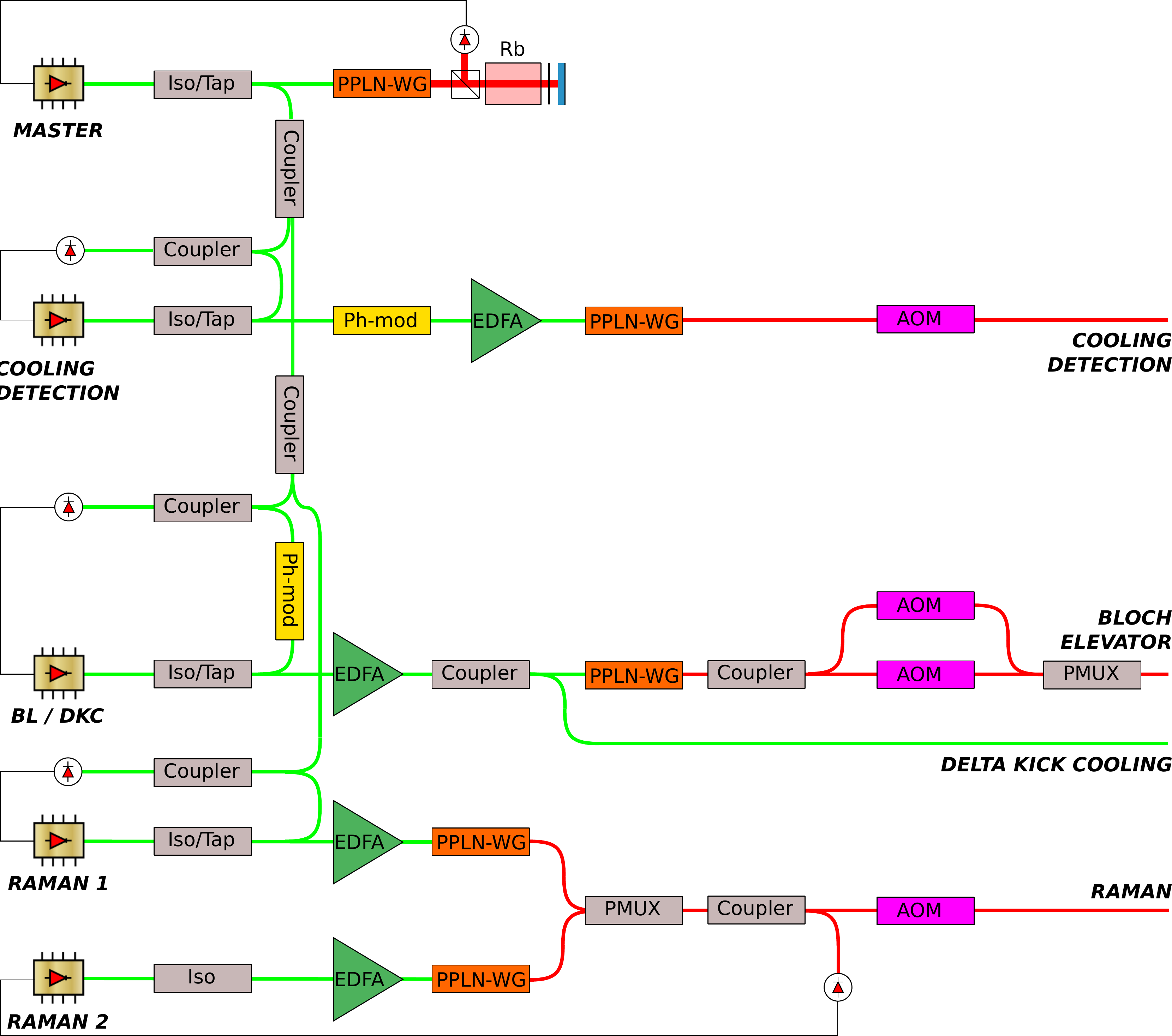}
}
\caption{Optical architecture of the laser system. Iso/Tap: optical isolator with tap coupler, PPLN-WG: waveguide PPLN crystal, Rb: Rubidium cell, Ph-mod: phase modulator, EDFA: Erbium-Doped Fiber Amplifier, AOM: Acousto-Optic Modulator, PMUX: polarization multiplexer.}
\label{fig:optical_architecture}       % Give a unique label
\end{figure}

The optical architecture of the ILS is based on a master-slave architecture (Fig. \ref{fig:optical_architecture}). A master laser is frequency-locked to an atomic transition with a known absolute frequency. Several slave lasers are in turn offset-locked on this master laser. Their frequency can be adjusted by tuning the setpoint of the lock. In the following paragraphs, we describe each of the functions of the laser system.

The master laser is a fibered laser diode emitting approximately 10~mW at 1560~nm (RIO Planex), current-modulated at 5~MHz. It is frequency-doubled in a waveguide PPLN crystal (NEL WH-0780) and the resulting 780~nm light is sent through a Rubidium spectroscopy cell in a double-pass configuration. The resulting saturated absorption spectroscopy signal is demodulated and used to lock the diode on the $|F=3\rangle \rightarrow |F'=3\rangle / |F=4\rangle$ crossover of $^{85}$Rb by retroacting on its current. A small fraction of the 1560~nm light is used to lock the slave lasers.

The cooling and detection laser is used to cool the atoms in a MOT followed by far-detuned optical molasses, and to detect them at the output of the interferometer. A RIO laser diode is offset-locked to the master laser with a tunable setpoint that can be adjusted in order to set the detuning of the laser with respect to the $|F=2\rangle \rightarrow |F'=3\rangle$ transition of $^{87}$Rb from 0 to 120~MHz. Light then goes through a phase modulator (Ixblue MPZ-LN-10) that generates sidebands at approximately 6.5~GHz, one of which being tuned on the $|F=1\rangle \rightarrow |F'=2\rangle$ transition for repumping. We then amplify the signal to approximately 1~W in a custom-made Erbium-Doped Fiber Amplifier (EDFA) with active output power stabilization, and frequency-double it in a fiber-coupled PPLN crystal. At this stage, the power at 780 nm is 500 mW. Finally, a custom-made fibered Acousto-Optic Modulator (AOM) module is used to set the output optical power. At the fiber output, the power is 260~mW.

The Bloch Elevator laser needs to be detuned from the master laser by approximately 100~GHz at 780~nm. We achieve this detuning by using a 1560 nm phase modulator driven with a 7~GHz RF signal with a power of approximately 25 dBm. We then select the 7th sideband to lock the laser with an offset of 49~GHz before frequency-doubling, resulting in a 98~GHz detuning at 780~nm. Initially, it was planned to realize the delta-kick collimation of the atom source using an optical dipole trap \cite{kovachy2015}, rather than the CAI mission atom chip itself \cite{muntinga2013,corgier2018}. This motivated the generation of a laser beam with a few watts of optical power, and our optical architecture combines this function with the Bloch Lattice. The 49~GHz detuned laser is amplified to 5~W in a high-power EDFA. 4~W would thus have been used for the dipole trap, and the remaining power is frequency-doubled and used for the Bloch elevator. This light is split in two paths and sent through two AOM modules that have adjustable frequencies, to create the required optical frequencies for the two counterpropagating Bloch lattices. The two beams are then recombined with orthogonal polarizations using a fibered polarization multiplexer. The output powers in the two beams are 100 and 156~mW.

Finally, we use two independant RIO laser diodes to drive the Raman transitions. With this configuration, we make sure that there are no spurious sidebands during the interferometer \cite{carraz2012}. The Raman 1 laser is offset-locked on the master laser with a Raman detuning of 3.4~GHz from the $|F'=1\rangle$ level to minimize spontaneous emission.  It is amplified to 1~W and frequency doubled. The Raman 2 laser is offset phase locked to the Raman 1 laser directly at 780~nm. This architecture is used to suppress any phase fluctuations due to physical path separation. The two Raman lasers are combined in the same fiber with orthogonal polarizations, and an AOM module is used to drive the Raman pulses. At the output, each Raman laser has a power of more than 400~mW.

The control electronics for the ILS were designed to minimize human supervision. All the lasers can be locked automatically through a dedicated software, which also checks that the laser remains stable on the long term. The software also improves the long-term stability of the lock by adding a digital integrator stage \cite{leveque2014}.

\subsection{Optical characterizations}
After integration, all the outputs of the ILS were tested individually in order to verify that they met the specifications required for the CAI mission. In this section we only describe the optical characterizations performed on the Raman output, which is the most critical of the system.

\begin{figure}
\resizebox{0.5\textwidth}{!}{%
  \includegraphics{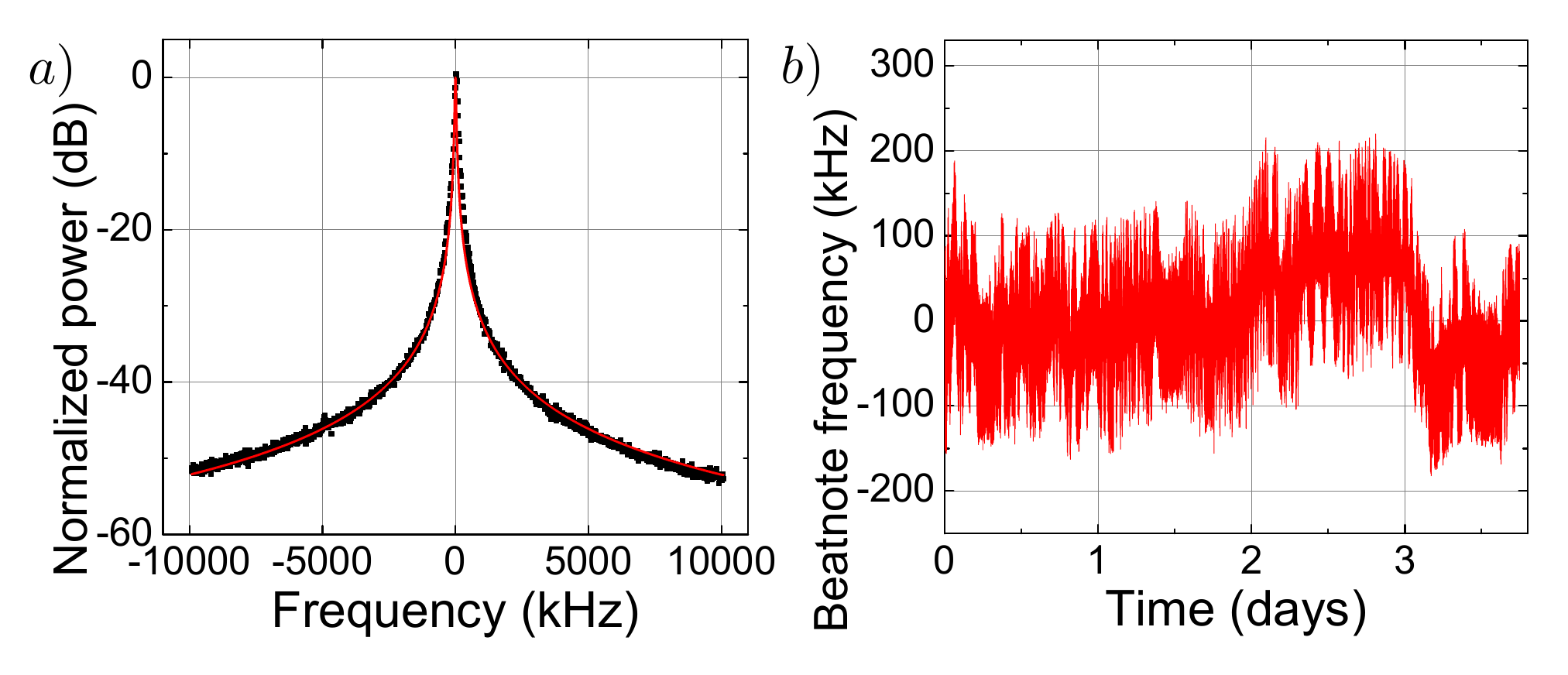}
}
\caption{Beatnote between Raman 1 laser and a similar independant laser. a) Beatnote recorded on a spectrum analyzer with a 9.1~kHz resolution bandwidth. A Lorentzian fit of the wings of the distributions gives a FWHM of 49~kHz. b) Center frequency of the beatnote recorded on a frequency counter. The standard deviation over more than 3.5 days is lower than 75~kHz.}
\label{fig:raman_linewidth}
\end{figure}

% Linewidth
The linewidth of the Raman 1 laser was measured by recording a beatnote between this laser and another fully independant system, built with similar specifications and locked using its own saturated absorption spectroscopy setup \cite{menoret2018}. The wings of the spectrum are fitted with a Lorentzian function, whose FWHM is the sum of the two individual widths. Preliminary characterizations have shown that the external laser used for the tests has a linewidth higher than 20~kHz. According to its specifications, the Raman 1 laser should have a linewidth lower than 10 kHz. The beatnote has a FWHM of 49~kHz (Fig. \ref{fig:raman_linewidth}a), which is compatible with the specifications of the two diodes.

% Frequency stability
By using the same setup and recording the center frequency of the beatnote over time, we evaluated that the long-term frequency stability of the laser was better than 75 kHz rms over several days (Fig. \ref{fig:raman_linewidth} b). This is well within the specifications, and shows that the laser system can remain locked over long periods of time without drifting, and without requiring human supervision. Since the Raman 1 laser is phase locked onto the master laser, these measured frequency fluctuations originate from the lock on saturated absorption peaks. The Raman 2 output was not tested with this setup. Indeed, being phase-locked to the Raman 1, it has the same frequency behaviour.

\begin{figure}
\resizebox{0.5\textwidth}{!}{%
  \includegraphics{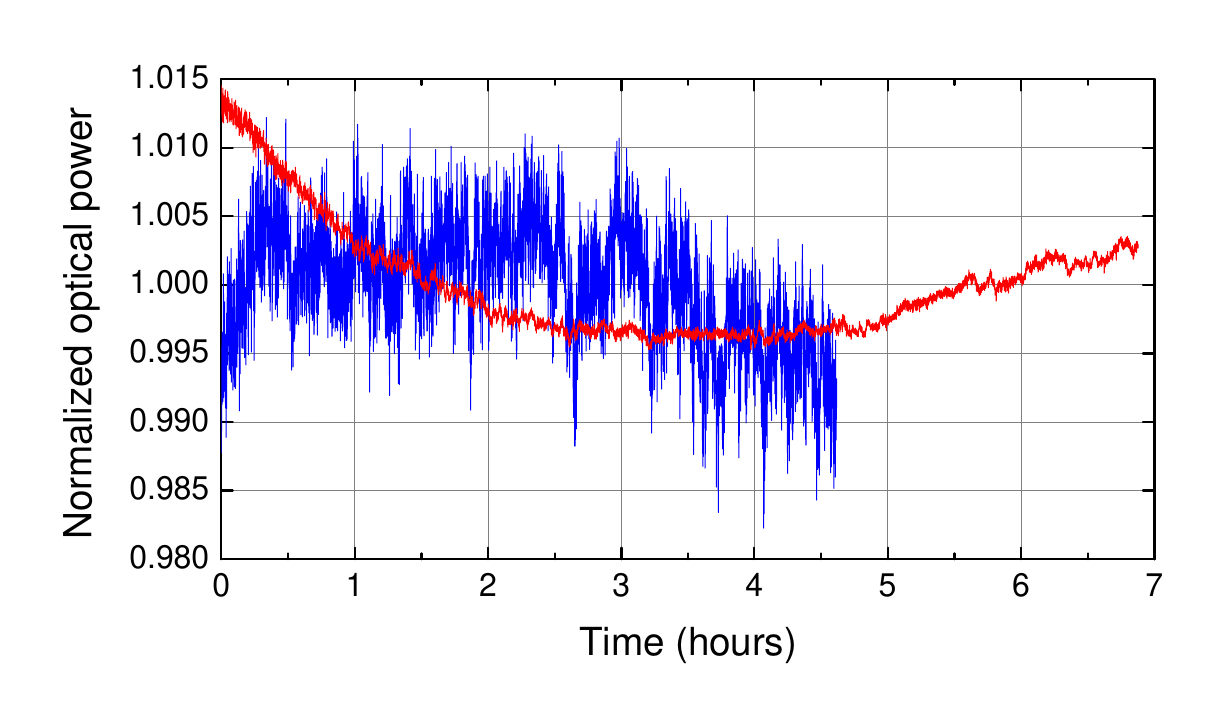}
}
\caption{Normalized power of the Raman 1 (red) and Raman 2 (blue) outputs over several hours. Average powers are 423~mW for Raman 1 and 415~mW for Raman 2. The two datasets were recorded at different times.}
\label{fig:raman_powerstab}
\end{figure}

% Power stability
Finally, we monitored the power in the two Raman outputs over several hours, to characterize the power stability of the system (Fig. \ref{fig:raman_powerstab}). Lasers were free-running during this measurement because the Raman 2 output can not be measured on its own if it is locked. The power of both lasers has a long term stability below 2\%. Raman 2 laser has a short-term noise (below 0.35\% rms in a 100~s window) which is higher than Raman 1 (0.05\% rms in 100~s). The two lasers being built with identical architectures, we attribute this difference either to a higher intrinsic noise in one of the EDFA pump diodes or to different settings in the servo-loop that controls the output power of the EDFAs. In both cases the noise is limited by residual low-frequency fluctuations and is compliant with the mission specifications: in a 2~s window, fluctuations are well below the level measured in 100~s.

Before using the ILS to cool and manipulate atoms, we have verified that all the other outputs have a similar behaviour to what we have presented here.

\section{Functional tests}

\subsection{Experimental setup}

The laser system was tested on a ground-based laboratory sensor under development at SYRTE. This sensor is designed for measurements of gravity and its gradient, based on simultaneous atom interferometers performed on two vertically separated atomic sources \cite{Caldani2019}. At the time of the test campaign, the setup was partly functional, with the bottom interferometer under operation only \cite{Langlois2017}. 

We only briefly describe this setup and the measurement sequence here, more details can be found in \cite{Langlois2017,Caldani2019}. At first, we trap about $10^8$ cold $^{87}$Rb atoms in 480~ms in a 3D mirror magneto-optical trap (MOT), which is loaded from the cold beam of a {2D MOT}. We then cool the atoms in a far detuned optical molasses before releasing them in free fall in the $|F=2\rangle$ hyperfine ground state and selecting them in the $|F=1, m_F=0\rangle$ state with a combination of microwave and pusher pulses. An atom interferometer is then realized with a sequence of three counterpropagating Raman pulses, equally separated in time, and aligned vertically. The interferometer phase shift is given by $k g T^2$, where $k$ is the effective Raman wavevector, $g$ is the gravitational acceleration and $T$ is the time interval between consecutive pulses. When the atoms are simply released after cooling, the duration of the interferometer $2T$ is limited to about 160~ms by the size of the vacuum chamber. After the Raman interrogation, the populations in the two output ports of the interferometer are detected using state selective fluorescence, out of which the interferometer phase is deduced.

To drive this sensor, we use the compact and simple home-made laser system (HMLS) described in \cite{Merlet2014}. This laser system, based on two semiconductor diode lasers and one amplifier, generates all the laser beams required for running a cold atom interferometer based on Raman transitions, namely laser cooling, interferometer and detection beams, and has been used for a number of different atom interferometer experiments \cite{Lautier2014,Trimeche2017,Langlois2017,Caldani2019}.

The availability of this well characterized HMLS allowed to validate individually each functionality of the ILS, and compare the performances obtained with the two systems. This was done by operating the sensor with the HMLS, with the laser beam of a given functionality exchanged for the corresponding one of the ILS. We report below the results of the validation and test campaign of the different functionalities.

\subsection{Detection}

We first tested the detection by exchanging the detection fiber of the HMLS with the cooling/detection output of the ILS. The frequency of the ILS laser was set so as to optimize the fluorescence signal, with the phase modulation turned off (as there is no need for repumping). We control the MOT parameters with the HMLS.

\begin{figure}
\resizebox{0.5\textwidth}{!}{%
  \includegraphics{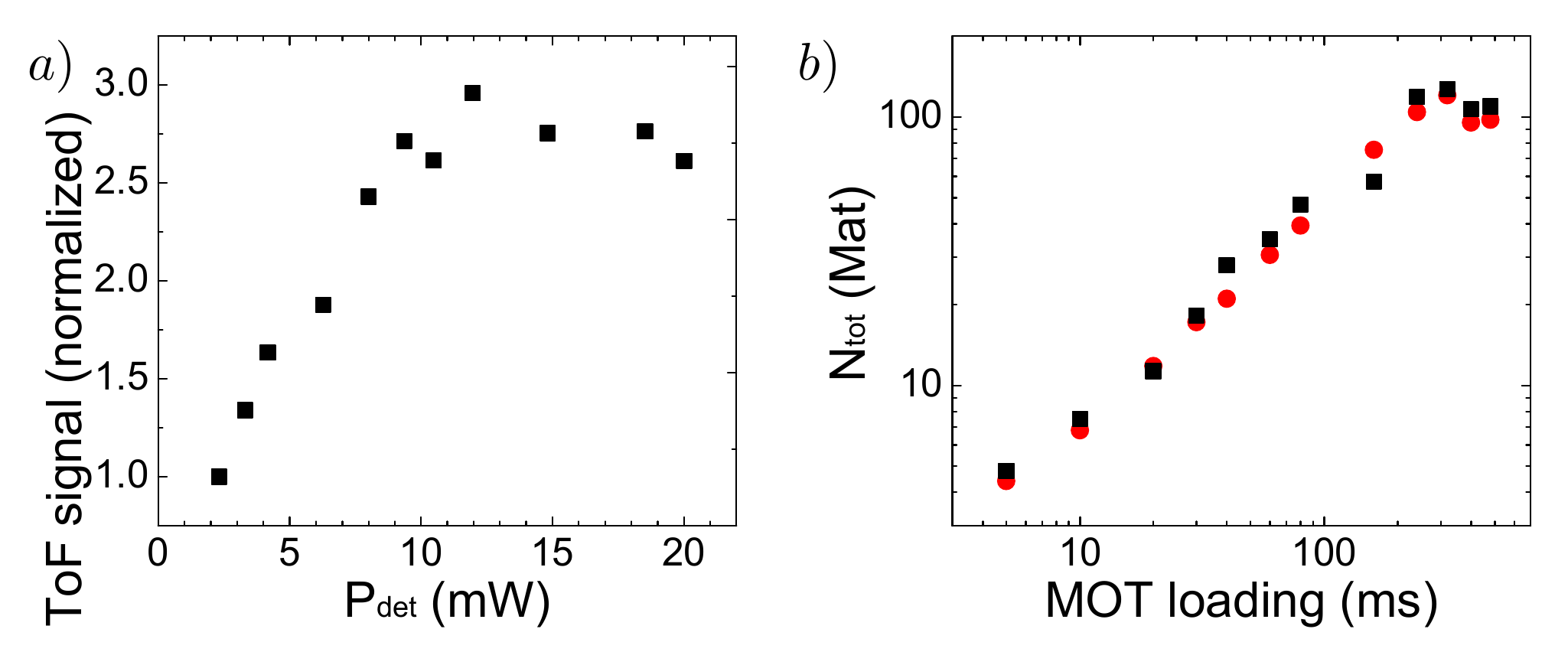}
}
\caption{Characterization of the detection laser. a) Detection signal obtained for different optical powers in the detection light sheets. b) Total atom number detected by the ILS (black squares) and HMLS (red circles) as a function of the MOT loading duration. The optical power in the detection light sheets is set to 2.3 mW.}
\label{fig:Det1}
\end{figure}

Figure~\ref{fig:Det1}a displays the fluorescence signal collected as a function of the optical power in the detection light beams. To compare this detection efficiency with the HMLS, we measured the total atom number detected with respect to the MOT loading time with both laser systems for the same optical power in the detection of 2.3 mW. We obtained very similar atom numbers as displayed in Fig.~\ref{fig:Det1}b.

\begin{figure}
\resizebox{0.5\textwidth}{!}{%
  \includegraphics{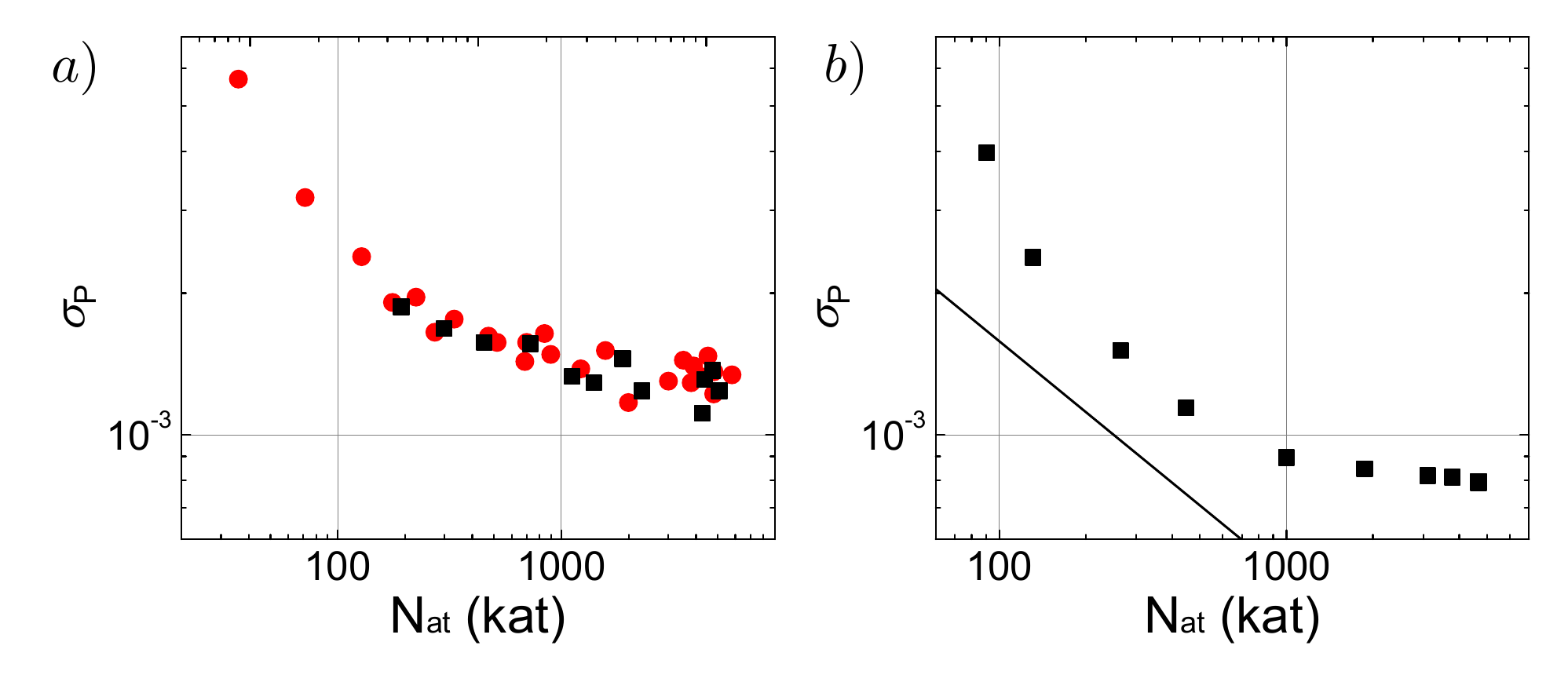}
}
\caption{Characterization of the detection noise. a) Allan deviations of the transition probability P as a function of the total number of atoms for identical optical power in the detection light sheets of 2.3 mW, for both the ILS (red circles) and HMLS (black squares). b) Same with the ILS only, with 4mW power. The solid line displays the quantum projection noise limit.}
\label{fig:Det2}
\end{figure}

In a second series of experiments, we measured the detection noise as a function of the detected atom number. For that purpose, atoms are prepared in an equal superposition of $|F=1, m_F=0\rangle$ and $|F=2, m_F=0\rangle$ states, using a combination of microwave and pusher pulses. Figure~\ref{fig:Det2} displays the Allan standard deviation of the measured transition probabilities $P$ as a function of the atom number. The atom number was varied by adjusting the parameters of the microwave selection sequence.

The results obtained with the nominal detection power of 2.3 mW are displayed on Fig.~\ref{fig:Det2}a and compared to the results we obtained with HMLS in the same conditions, showing identical behaviours, suggesting that noise contributions arising from intensity and frequency fluctuations of the lasers do not play a major role here. 

Such measurements were performed at different detection powers of the ILS. Figure~\ref{fig:Det2}b displays the results obtained with a detection power of 4 mW, for which the noise lies less than a factor of two above the quantum projection noise limit, for atom numbers in the $10^5 - 10^6$ range. We actually found that for larger laser powers, the noise on the transition probability does decrease at low atom number, as expected, but degrades at high number of atoms.

\subsection{MOT}

To evaluate the laser cooling functionality, we connected the same cooling fiber from the ILS to the MOT fiber splitter which distributes the beams to the 2D and 3D MOTs. Here the amplitude of phase modulation is adjusted such as to put a few percents of power in resonance with the repumping line. The computer control system of the sensor was used to trigger the DDS frequency sysnthesizer that controls the frequencies (cooling and repumper) of the ILS cooling laser, so as to modify the frequency of the cooling light during the cycle synchronously with the rest of the sequence.

\begin{figure}
\resizebox{0.5\textwidth}{!}{%
  \includegraphics{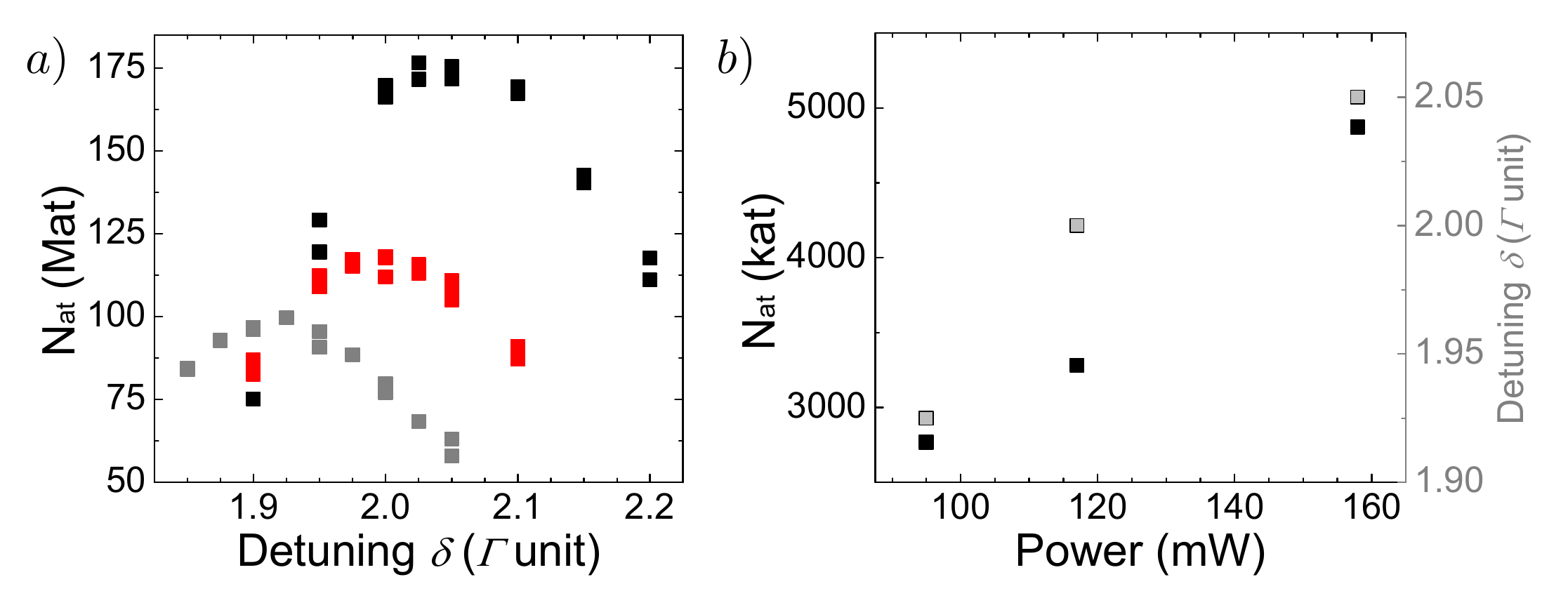}
}
\caption{a) Number of trapped atoms as a function of the detuning, expressed in units of $\mathrm{\Gamma}$, for three different laser cooling powers : 95mW (grey), 117mW (red) and 158mW (black). b) Optimal number of atoms (black) and corresponding optimal detunings (grey) as a function the optical power.}
\label{fig:Mot}
\end{figure}

Figure~\ref{fig:Mot} represents the number of atoms detected as a function of the detuning of the cooling line for three different cooling optical powers, after a loading time of 480 ms. For the same MOT parameters (frequency and optical power corresponding to the grey squares in Fig. \ref{fig:Mot} a) that we use on the SYRTE experiment with HMLS, we obtained here the same number of atoms, and twice as much for an optical power of 158~mW. As expected, the optimal detuning is found to increase with the laser power.

\begin{figure}
\resizebox{0.5\textwidth}{!}{%
  \includegraphics{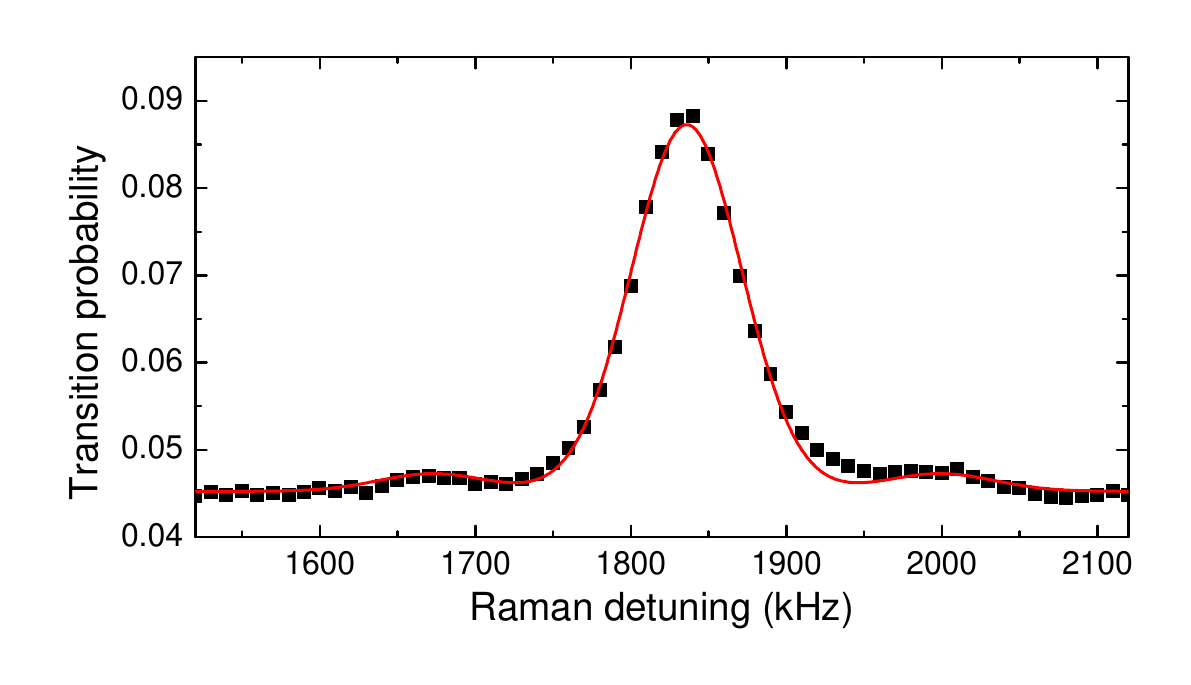}
}
\caption{Raman spectrum measured after the molasses sequence. The Gaussian fit of the main peak gives a $1/e^2$ width of 70 kHz, corresponding to a molasses temperature of 2 $\mathrm{\mu}$K.}
\label{fig:Temps}
\end{figure}

To validate the capability of the ILS to cool the atoms in the far detuned molasses phase down to the temperature limit of sub-Doppler cooling, we measured the temperature of the atoms after their release from the cooling beams using Raman velocimetry. Using the Raman functionnality of the HMLS, we applied a 80 $\mathrm{\mu s}$ long Raman $\pi$ pulse after a free fall delay of about 70 ms, and recorded the transition probability as a function of the Raman frequency  (Fig.~\ref{fig:Temps}). The spectrum then corresponds to the distribution of Doppler shifts, and thus to the atomic velocity distribution. One observes as well two side resonances corresponding to parasitic Zeeman shifted conterpropagating transitions. The $1/e^2$ width of the spectrum of 70 kHz corresponds to a temperature as low as 2 $\mathrm{\mu}$K.

\subsection{Bloch elevator}
Next, we connected the Bloch elevator fiber output of the ILS to the Raman collimator instead of the Raman beam fiber, in order to generate a Bloch lattice in the vertical direction. In addition, a polarization beamsplitting cube was inserted before the retroreflecting optics in order to suppress one of the two Bloch beams before the retroreflection so as to obtain one lattice only, instead of two lattices moving in opposite directions as proposed in \cite{trimeche2019}. 

\begin{figure}
\resizebox{0.5\textwidth}{!}{%
  \includegraphics{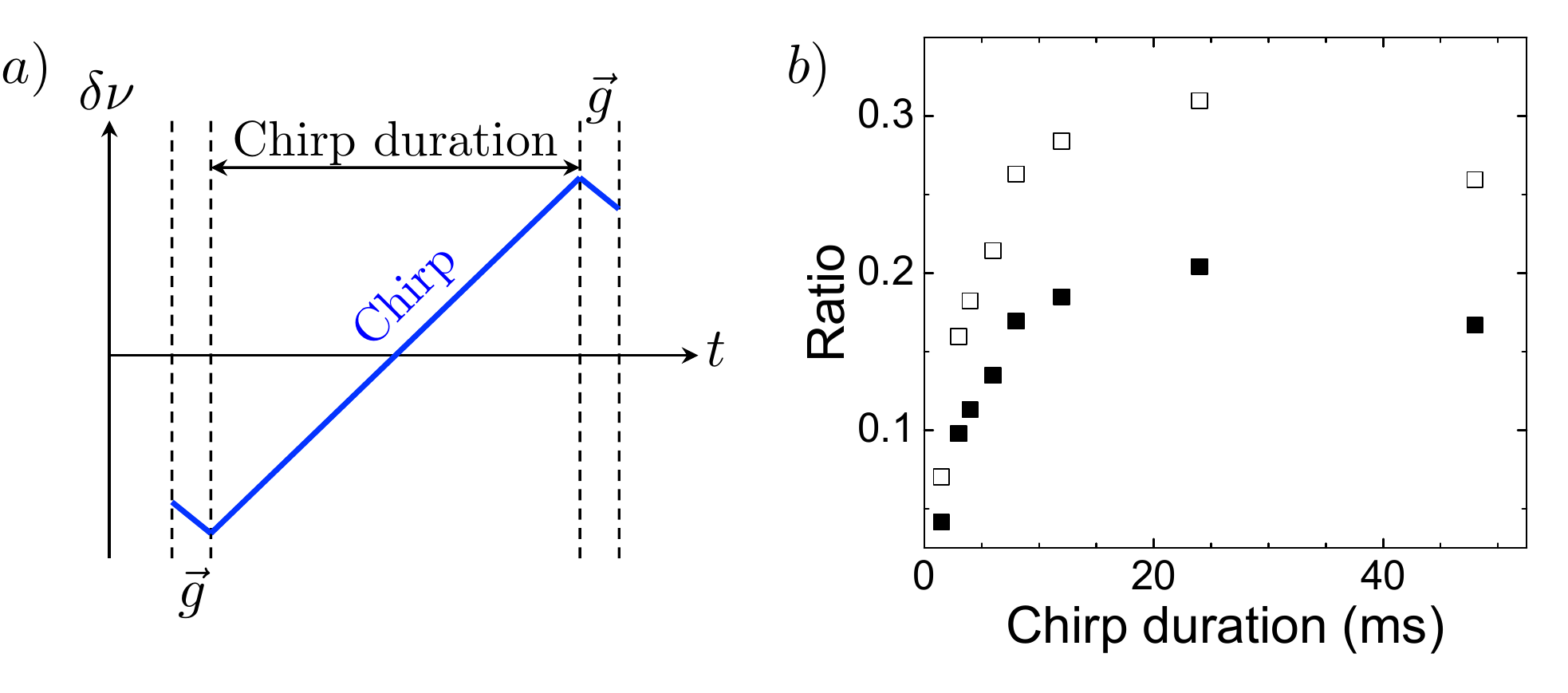}
}
\caption{Launching the atoms using a Bloch elevator. a) Frequency difference between the two optical beams during the launch. The adiabatic loading and release phases are 200$\mu$s long. In between, the length of the acceleration phase can be modified. b) Efficiency of the launch with BO for different acceleration phase durations. The direct measurements of the fraction of launched atoms are displayed as filled black squares. These ratios are then corrected to account for the change in the velocity of the atoms in the detection sheets and in the cloud expansion. Corrected data are displayed as empty squares.}
\label{fig:Bloch}
\end{figure}

The frequencies of the lattice beams are controlled by adjusting the frequencies of the ILS AOMs using programmable DDSs. This leads to the sequence displayed on Fig.~\ref{fig:Bloch}a: first, the adiabatic loading of the lattice, for 200 $\mu$s, where the frequency difference between the Bloch beams matches gravity acceleration, second, the launch phase, where the chirp on the frequency difference is reversed and much increased in order to accelerate the atoms upwards and third, the adiabatic release where the frequency sweep matches gravity acceleration.

We measured the fraction of launched atoms as a function of the duration of the acceleration phase for a fixed number of 100 Bloch Oscillations (BO). Figure~\ref{fig:Bloch}b displays the results as a function of the chirp duration. The best result is obtained for a chirp of 24 ms, which corresponds to a modest acceleration of 5~$g$.

The measurements shown in Fig.~\ref{fig:Bloch}b are corrected in order to account for the finite size of the detection sheets and the difference between launched and dropped atoms in terms of velocity at the detection and ballistic expansion. Indeed, the faster the atoms, the shorter the time they spend in the detection light sheets and the smaller the fluorescence signal.

These results are comparable in terms of fraction of launched atoms to the ones we obtained in~\cite{Langlois2017}. This fraction corresponds more or less to the fraction of the velocity distribution fitting into the first Brillouin zone (of width $\mathrm{2v_r}$). In~\cite{Langlois2017}, we obtained a corrected ratio of up to 40 percent for 1.8 $\mathrm{m.s^{-1}}$ launch velocity, and an acceleration duration of 2 ms, owing to the higher laser power (300 mW per beam instead of 100 mW) and smaller detuning (50 GHz instead 100 GHz).

\subsection{Atomic Interferometer}

We finally connected the Raman fiber of the ILS to the Raman collimator. A frequency chirp $\alpha$ generated by one of SYRTE's DDS is applied to the phase lock loop of the ILS Raman lasers. The output optical power of each of the two Raman beams was controled independently by adjusting the output level of the EDFAs with the ILS control software. The total power of the Raman output was controled by changing the RF power on the Raman AOM, using a controlled voltage provided by SYRTE’s control system. We modified the experiment in order to optimise the performances for continuous gravity measurements. Indeed, as the phase difference between the Raman lasers is linked to the position of the mirror, fluctuations of this position due to ground vibrations can induce significant interferometer phase noise, washing out the interferometer fringes. We thus used a passive isolation plateform and recorded the remaining vibration noise with a low-noise seismometer (Guralp 40T). We then post-corrected the interferometer phase from the effect of these residual vibrations, which improves the sensitivity of the measurement~\cite{LeGouet2008,Merlet2009}. In order to measure the mirror vibrations as accurately as possible, this mirror is directly fixed on the seismometer that we placed at the top of the experiment. The distance from the atomic sample's initial position to the detection sheets allowed for performing interferometers with durations of up to $2T=160$~ms. The total cycling time was 460~ms.

\begin{figure}
\resizebox{0.5\textwidth}{!}{%
  \includegraphics{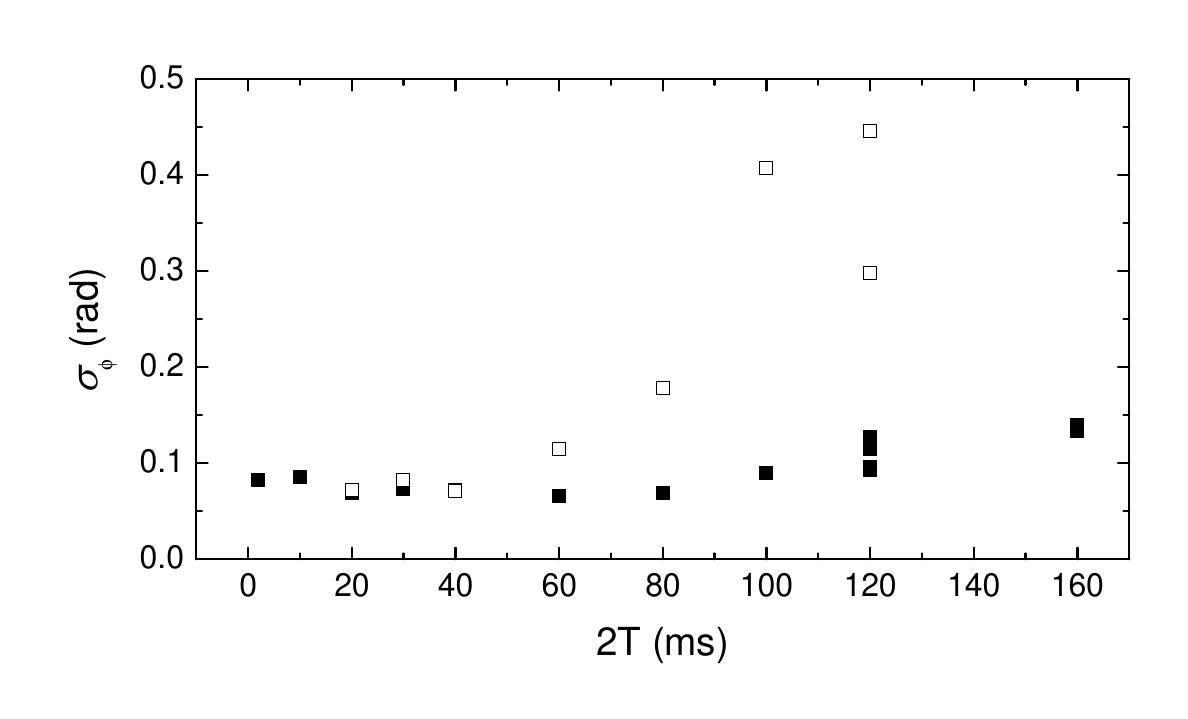}
}
\caption{Interferometer phase noise as a function of the interferometer duration $2T$, with (black squares) and without (open squares) vibration correction.}
\label{fig:noise}
\end{figure}

We first measured the short term phase noise of the interferometer as a function of the interferometer duration $2T$. Results are presented on Fig.~\ref{fig:noise}, where the measured noise with and without the correction from vibration noise is displayed with black filled and empty squares respectively. We observe as expected a significant increase of the phase noise with $T$ due to the increasing impact of the vibration noise. This increase is efficiently mitigated by the correction. Also, we find an asymptotic level of noise at low $2T$, ie free from the impact of vibration noise, of order of 50-100 mrad per shot. This is well above the typical level of detection noise, which for our parameters, contributes up to $\sigma_{\Phi}=2 \sigma_{P,det}/C = 8$ mrad per shot, with $\sigma_{P,det}=10^{-3}$ the detection noise on the transition probability and $C$ the contrast of the interferometer ($C$=26\% for $2T=160$~ms).

To investigate the source of excess noise, we also performed AI measurements with co-propagating Raman beams, and for different interferometer durations $2T$ and Raman pulse  durations $\tau$, and compared the phase noise for co- and counter-propagating Raman beams.

\begin{table}
\caption{Interferometer phase noise measurements for different AI measurement conditions.}
\label{Tab}       % Give a unique label
% For LaTeX tables use
\begin{tabular}{lccccc}
\hline\noalign{\smallskip}
Configuration & $2T$&$\tau(\pi /2$) & $C$ & $\sigma_\mathrm{P}$ & $\sigma_\mathrm{\Phi}$/shot \\ 
• & (ms) & ($\mu$s) & \% & ($\times 10^{-2}$) & (mrad) \\
\noalign{\smallskip}\hline\noalign{\smallskip}
Counter-prop & 160 & 4 & 26 & $1.6$ & 123 \\
Co-prop & 160 & 4 & 75 & $1.8$ & 48 \\
Co-prop & 2 & 4 & 90 & $2.6$ & 57 \\
Co-prop & 2 & 80 & 68 & $6.1$ & 17 \\
\noalign{\smallskip}\hline
\end{tabular}
\end{table}

Table~\ref{Tab} presents some of these sensitivity measurements. The most relevant set of parameters for gravity measurements is the first one: it results in a sensitivity of $8.10^{-8}g$ at 1s (82.6 $\mathrm{\mu}$Gal), close to the one obtained in a compact gravimeter based on a similar laser architecture \cite{menoret2018}. It is about one order of magnitude higher than the best sensitivity obtained with the state of the art SYRTE Cold Atom Gravimeter (CAG)~\cite{Gillot2014} which operates with a shorter cycling time (380ms), a better isolation plateform, a more linear and lower noise seismometer and with an anti-acoustic enclosure.

The phase sensitivity in co-propagating measurements is more than twice better for the same duration of Raman pulses ($\mathrm{\tau_{\pi/2}}$), irrespective of the interferometer duration $2T$. This sensitivity improves when increasing the duration of the Raman laser pulses. This indicates that the noise is dominated by the contribution of the laser phase noise at high Fourier frequency. These results are compatible with the expected phase noise, calculated out of the power spectral density of Raman phase fluctuations. This is not a fundamental limitation of the system, phase noise can be improved by optimizing the electronics and by using a longer pulse duration.

\begin{figure}
\resizebox{0.5\textwidth}{!}{%
  \includegraphics{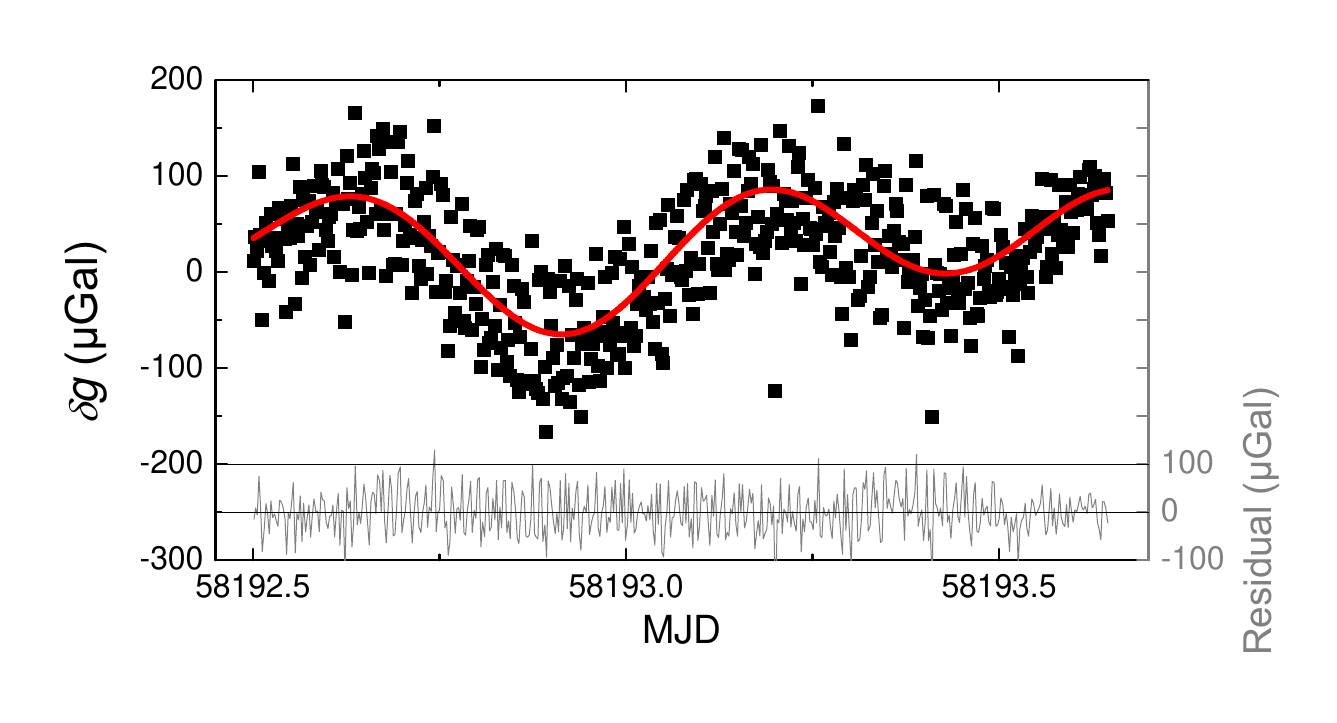}
}
\caption{Gravity measurements (black dots), uncorrected from tidal effects (red line). The data are averaged over 184s. The difference is represented in grey.}
\label{fig:g}
\end{figure}

\begin{figure}
\resizebox{0.5\textwidth}{!}{%
  \includegraphics{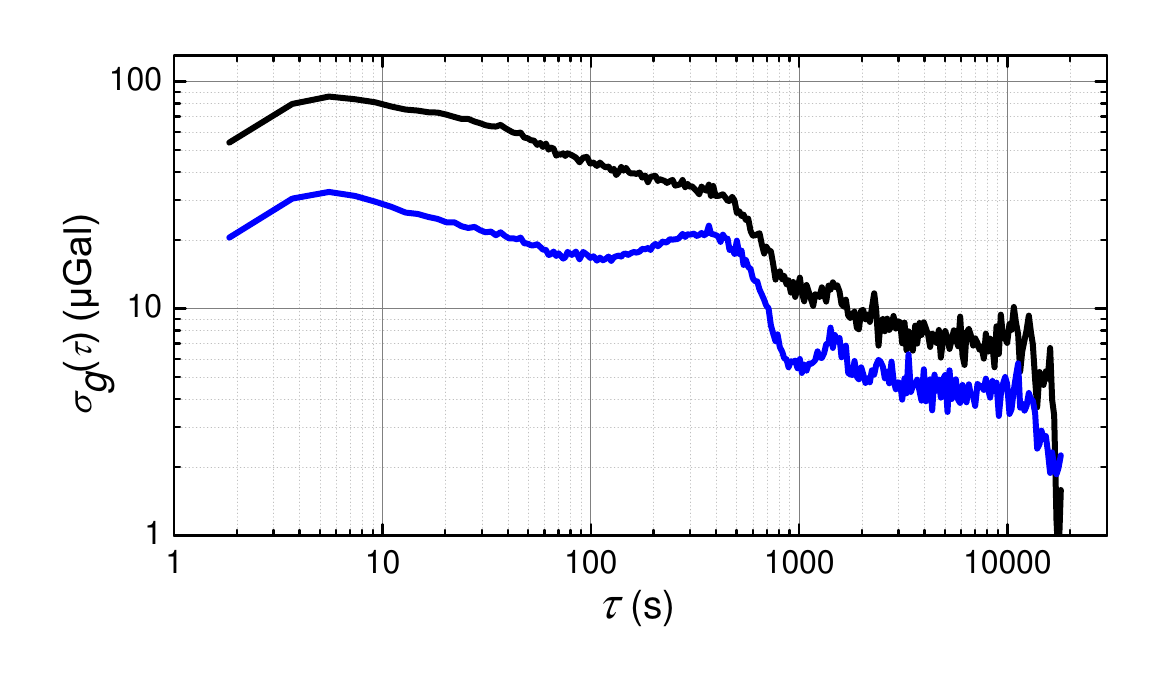}
}
\caption{Allan deviations of gravity fluctuations. The black line corresponds to the residuals of Fig. 10, which are corrected from tides and two-photon light shift. The blue line displays the average over the four measurement configurations, which is also corrected from tides, but not from two-photon light shift.}
\label{fig:var}
\end{figure}

We performed gravity measurements using the four configurations measurement protocol we routinely implement for the CAG measurement to reject most of the systematic effects which affect the gravity measurement. This protocol, based on interleaved measurements with opposite Raman wavevector orientations $k_\uparrow$ and $k_\downarrow$ and two different Raman laser intensities, is described in detail in~\cite{Louchet2011}. The results, displayed in Fig. \ref{fig:g}, follow the expected gravity variations due to tides. The difference with the local tide model is represented in grey. The Allan standard deviation of this difference is displayed as a black line in Fig. \ref{fig:var}. As discussed in \cite{Louchet2011}, the correction of the two-photon light shift leads to a degradation of the short-term sensitivity with respect to, for instance, an average over the four measurement configurations simply corrected from tides, which is displayed as a blue line. Here, the Raman power of the ILS is sufficiently stable, meaning that fluctuations of two-photon light shifts are not a limitation to the long-term stability of the measurement. Therefore, using the complete protocol alternating over four configurations is not necessary in this case. A basic alternation of $k_\uparrow$ and $k_\downarrow$ is sufficient for measurements that are both stable and precise, provided that the two-photon light shift has been previously measured. This would reduce the effective cycling time and improve the short term sensitivity.

\section{Conclusion}
We have described a prototype industrial laser system designed for atom interferometry measurements, in the perspective of future space missions with quantum inertial sensors. We identified frequency doubled telecom lasers as the most mature technological solution. It benefits from the availability of a wide range of components from several suppliers, most of which are qualified to the Telcordia standard. With this study, we reach a Technology Readiness Level (TRL) of 4, corresponding to the validation of a propotype in a laboratory environment \cite{esa2008}.

The prototype industrial system presented here was fully characterized and all of its functions were tested quantitatively on a ground-based cold atom interferometer. The system was found to comply with all of the mission specifications, and had performances comparable to that of a home-made laboratory laser system. Moreover, the system was operated during several weeks without requiring human supervision, and used to study differential phase extraction in the gradiometer experiment \cite{Caldani2019}.

Most of the components used in this prototype have been validated individually to a higher TRL, or have a commercial alternative available which is qualified \cite{leveque2014}. The only major exception are the PPLN waveguide modules used for frequency doubling. These devices will require additionnal hardening before they can be used in space. A way towards this goal is the use of micro-optical assembly techniques to package the crystal in a device that is compatible with space environment.

Further improvement of the TRL will require testing in environments representative of a space mission. These include vibrations, shocks, radiations, thermal cycling and operation under vacuum. We believe that meeting environment requirements can be achieved with a system similar to the one presented here with only minor adaptations, for example regarding the integration of the frequency doubling modules.

This work is a first step towards a space-qualified laser system for cold atom interferometry missions. We have validated the technological choice with a fully functional optical architecture in the frame of a case study for a gravity gradiometry mission. The optical architecture can easily be adapted to meet the requirements of different mission scenarios, for example in the field of fundamental physics \cite{hogan2011,altschul2015,Ertmer2009}.

\section*{Acknowledgements}
This work is supported by contract 4000116740/16/NL/MP from the European Space Agency. R.C. thanks the support from LABEX Cluster of Excellence FIRST-TF(ANR-10-LABX-48-01), within the Program Investissements d'Avenir operated by the French National Research Agency (ANR). The authors acknowledge a major contribution from the Muquans team in the integration and optimization of the ILS.

\section*{Author contributions}
The ILS was designed by VM, GS, BD, SM and FPDS. ASM, GS and VM (resp. RC, SM and FPDS) performed the optical (resp. functional) characterizations. All the authors contributed to the data analysis and manuscript preparation. All the authors have read and approved the final manuscript.

\bibliographystyle{epj}
\bibliography{biblio}

\end{document}